\documentclass{mem}
\usepackage{natbib}\usepackage{txfonts}\usepackage{balance}
\usepackage{graphicx}
\usepackage[a4paper]{hyperref}
\idline{75}{282}
\begin{document}
\def\teff{$T\rm_{eff }$}
\def\kms{$\mathrm {km s}^{-1}$}

\title{
Young stellar populations in the Magellanic Clouds
}

   \subtitle{}

\author{
A. \,Vallenari\inst{1} 
          }

\offprints{A. Vallenari}

\institute{
Istituto Nazionale di Astrofisica --
Osservatorio Astronomico di Padova, Vicolo Osservatorio 5,
I-35122 Padova, Italy
\email{antonella.vallenari@oapd.inaf.it}
}

\authorrunning{Vallenari }

\titlerunning{Young population in the MCs}

\abstract{We discuss the young population of stars and clusters in the Magellanic Clouds. We present the discovery of pre-main sequence candidates in the nebula N~11 in the Large Magellanic Clouds. The comparison of the Colour-Magnitude diagram with pre-main sequence tracks and the presence of Spitzer objects YSO I and II suggest that the star formation has been active for a long period in the region, from  a few $10^5$ yrs  to several Myr ago}



\maketitle{}

\section{Introduction}
The process of star formation in different environments is far from
being understood. In particular, it is difficult to reconcile the
prominent influence of the local environment (turbulence, compression,
initial trigger) on small scales with the universality of the Schmidt
and Kennicut law on Galactic scales which suggests that Galactic-scale
gravity is involved in the first stages of star formation.
Detailed understanding of how an entire molecular cloud converts into stars is still lacking. 
To analyze  star forming regions can cast light on the formation scenarios, since  spatial distributions of different components (molecular clouds, ionized gas), and of stars in different evolutionary stages can help to distinguish between the proposed models. Local Group galaxies, and in particular the Magellanic Clouds (MCs), are very promising regions to study the process of star formation. Their young population of stars and clusters allow to derive hints about the formation process in low metal content environments. This is particularly important since the formation of stars depends on the balance between the gas heating and the cooling, on which  the presence of metals has a significant effect.  
In this paper, in Section~\ref{forma} we first discuss the formation mechanism of field stars and clusters, in section~\ref{trigger} the  effects on the star formation of the gravitational interaction between the MCs and the Milky Way are discussed, 
in section~\ref{n11a} and \ref{n11b} the discovery of pre-main sequence candidates in N~11 in the LMC is presented. 
Final remarks are  given in section \ref{conclu}.

\section{Field star and cluster formation scenario}
\label{forma}
Several mechanisms of cluster and field star formation have been proposed in literature. In the  turbulence theory, star and cluster formation are regulated by the balance between turbulence and  gravitation.  
The net  effect of highly compressible turbulence is to prevent collapse globally, while on local scale it causes local density enhancements that might produce a collapse, under suitable conditions. If the surrounding flow is not strong enough to continue to drive the cloud, the turbulence will quickly dissipate giving rise to an active star formation. The properties of the fragmenting clouds and the mass of the individual proto-stellar cores, giving birth to  isolated stars rather then clusters of objects,  depend in a complex and still unclear way on the values of the politropic index which, in turn, is related to the metal content of the gas. When turbulence dominates, the star formation is inefficient and slow and stars build up in small groups.  
Turbulent control of the star formation predicts that star clusters form in regions where the support of the turbulence is insufficient  or where only large scale driving is working (McLow \& Klessen 2004).
Star formation on scales of galaxies as a whole is expected to be controlled by the balance between turbulence and self-gravity just like star formation on scale of individual gas clouds, but might be modulated by additional effects, such as cooling, differential rotation, gravitational interaction (Sasao 1973, Li et al 2004). 
An  alternative mechanism is suggested by Bekki et al (2004): during galaxy interactions and mergers  clusters  can  form  as a result of relatively high velocity cloud-cloud collisions.

\section{Young Cluster and field SF in the SMC: star formation by gravitational trigger}
\label{trigger}
Star formation by gravitational trigger is well known to take place in mergers/ interactions. The Magellanic Clouds represents one of the best studied examples. 

Formation episodes involving both cluster and field star formation happened at 5, 20, and finally at 100-150  Myr, in coincidence with  {SMC} perigalactic passage (Chiosi et al 2006).
As far as the old population is concerned,  the star formation
was not continuous but proceeded in a number of bursts taking place
at  3 Gyr and  6 Gyr ago.
These bursts  are  temporally coincident with past
peri-galactic passages of the SMC about the Milky Way.
Only a very low efficiency of the star formation
at epochs older than 6-8 Gyr is found (Chiosi \& Vallenari 2007, Tosi et al 2007, Noel et al 2007).
These  results point out  that the  most recent interactions between  {SMC} and  {LMC}  triggered cluster and field star formation in the  {SMC}, in agreement with the expectations from Bekki \& Chiba (2005) while at older ages, the tidal interaction
between the Magellanic Clouds and the Milky Way was not able to
give rise to significant star formation events (see Vallenari 2007 for a wider discussion).  
However, as expected, the star formation process cannot be explained uniquely as the result of gravitational trigger. Comparing field stars  with the cluster age distribution,
Chiosi et al (2006) find that there is not a complete coincidence  between young cluster and field star formation, suggesting  that different modes of formation might be at work.

\begin{figure*}[ht]
\centerline{
\resizebox{7.5cm}{!}{\includegraphics[clip=true]{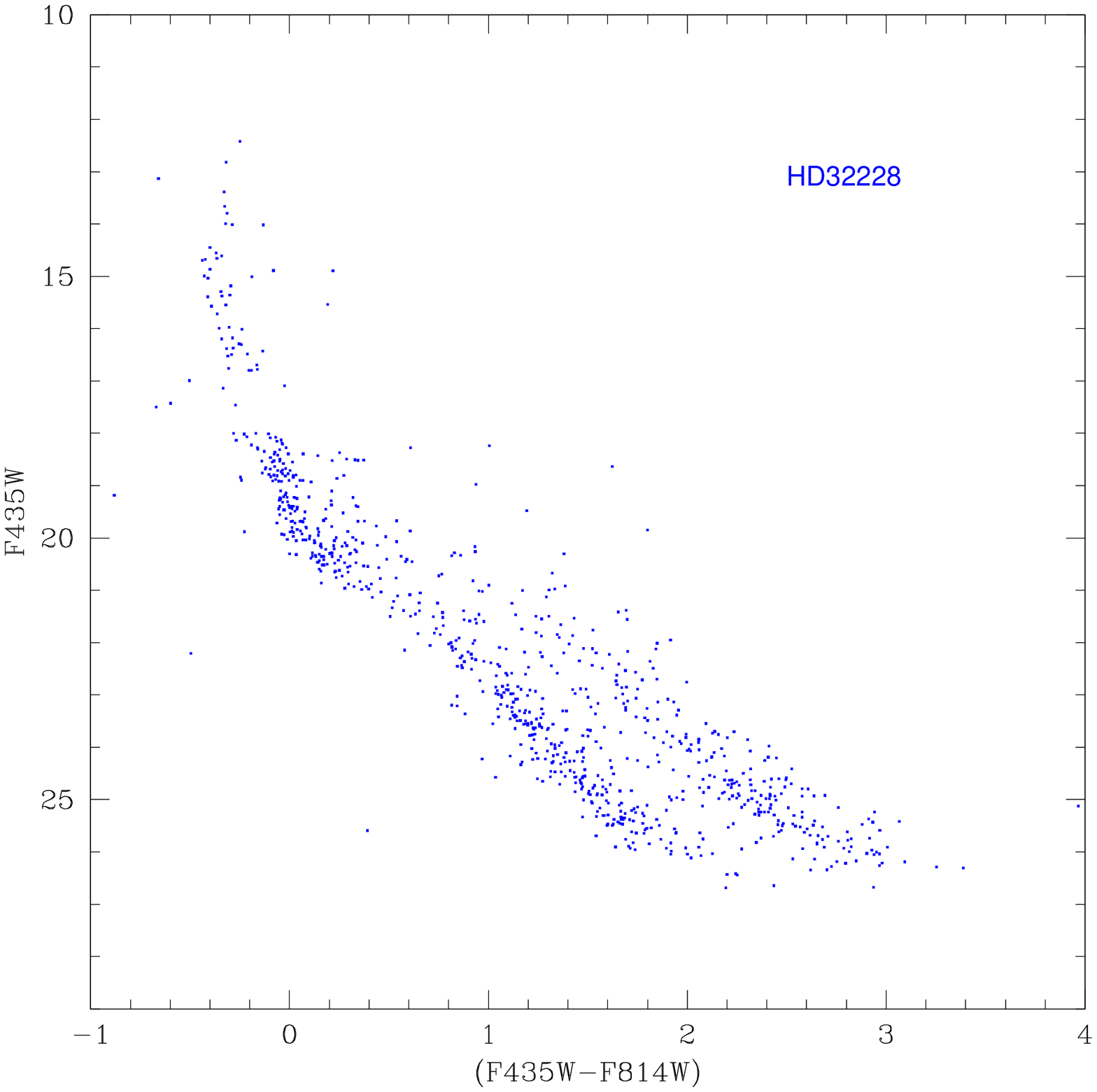}}~
\resizebox{7.5cm}{!}{\includegraphics[clip=true]{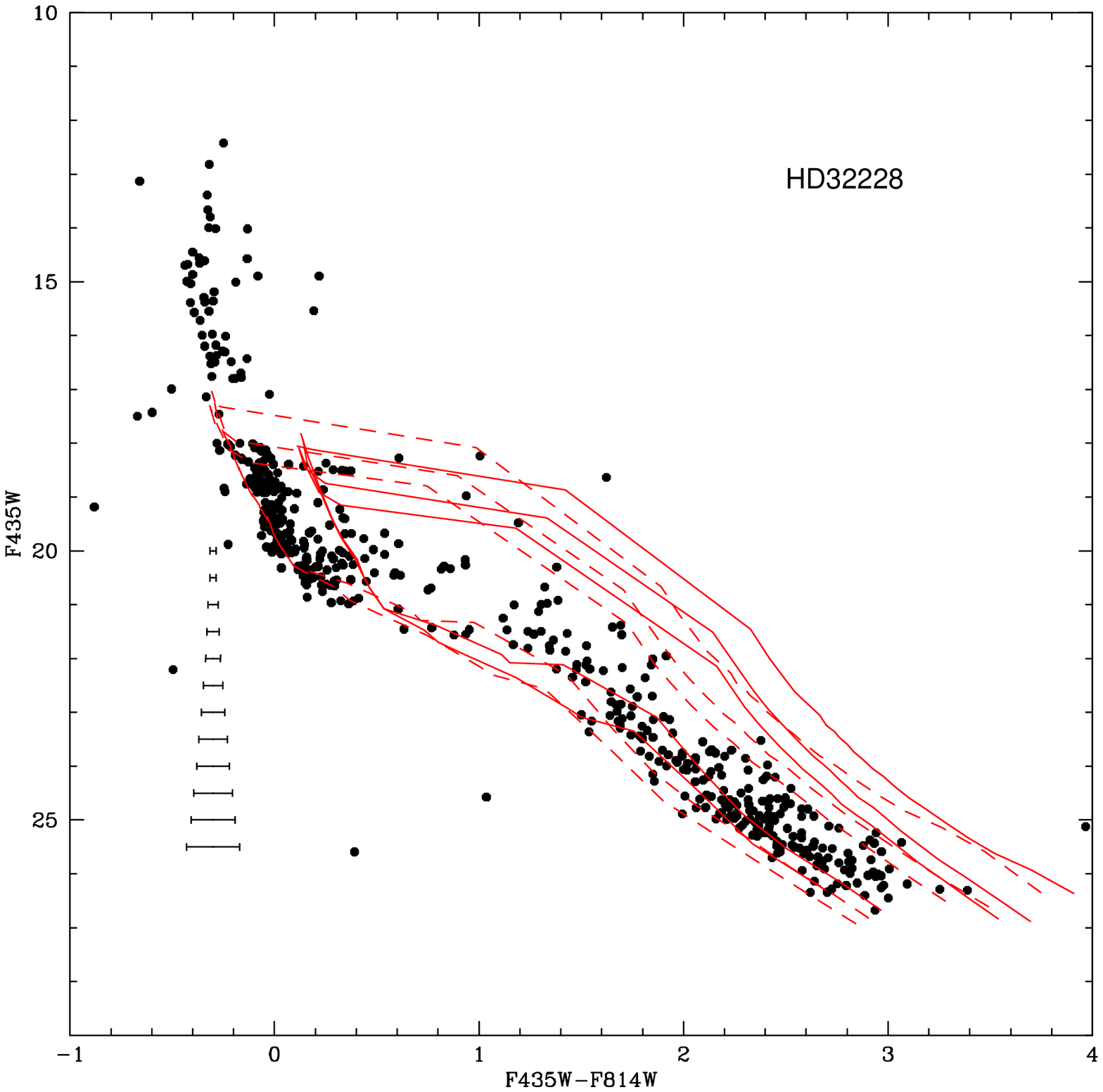}}~
}
\caption{
\footnotesize
HD~32228 is the central cluster of the association LH~9. Its age is derived by
Walborn et al (1999) on the basis of WC star found there. The PMS sequence is clearly 
visible and well separated by the MS stars (left panel).
Right panel: CMD of HD~3228 after that the field population contamination is subtracted.  Pre-main sequence isochrones by Siess et al (2000) are plotted for  A$_V=0.2$ (dashed lines), A$_V=0.8$ (solid lines) and ages of 0.5, 1.0,1.5,10, 15 Myr. Photometric errors on the color are shown for increasing values of the magnitude (solid error bars).
}
\label{li_vhel}
\end{figure*}

\section{The stellar population of N~11 in the LMC: a case of triggered star formation?}
 \label{n11a}
Most stars form in groups and clusters. This implies that recently formed stellar objects must interact with their environment. In this respect, one of the most intriguing aspects of the star formation is that of the triggering.
Young massive stars are expected to inject energy into the nearby interstellar medium, heating and compressing the surrounding gas.  This can have destructive or constructive effects, depending on the balance between heating and gravity but it is still unclear what regulates it.
To distinguish between sequential  and triggered star formation is indeed very difficult.  Star formation is often found to be sequential, i.e. stellar population can independently form in adjacent clouds with little age difference, but no causal relationship is present. A hierarchical system of three or more generations of stars in the same region is often interpreted as more stringent sign of triggers (i.e.  causal relationship between episodes of star formation)(Oey et al 2005, Deharveng et al 2005).
 The stellar population in N~11  is often presented as one of the best example of triggered star formation.
N~11 is the second largest nebula of the LMC after the 30~Dor
Nebula. It is located in the north-western corner of the LMC.  It has a peculiar morphology with a central hole with no emissions
and several filaments around. 
 The central cavity has been evacuated by the OB stars in the association LH~9, located near the center of N~11. The
whole complex has a diameter of 45', corresponding to a linear
extent of 705 pc, assuming a distance for the LMC of 54 kpc. In addition to LH~9, this complex includes the OB associations LH~10 in N11B, LH~13 in N11C, LH~14 in N11E.
Bica \& Dutra (2000) find at least 18 objects, including associations and clusters, in the age range 0-30 Myr.

Walborn \& Parker (1992) suggest a two stage star-burst hypothesis: an initial star burst in LH~9 triggered a secondary burst in LH~10, a few Myr later.
This hypothesis was recently confirmed by the results by Hatano et al (2006), Mokiem et al (2007)
 who detect 127 Herbig Ae/Be star candidates  from near-infrared photometry in this region, mainly in the periphery of LH~9. Herbig Ae/Be star are intermediate mass pre-main sequence stars (7-3 M$_\odot$) and have age range 1-3 Myr.
 Hatano et al (2006)  find a spatial correlation of OB stars and Herbig Ae/Be star candidates with the radio continuum, H$_\alpha$, CO and X-ray.  Herbig Ae/Be stars are found in all the associations, but inside LH~9. This suggests that in LH~9 such very young objects would have already disappeared, losing their circumstellar envelopes/disks. This leads to the conclusion that LH~9 is slightly older than the other associations and has possibly triggered star formation episodes in the surrounding regions.

\section{Pre-MS candidate stars in N11}
\label{n11b}

PMS stars are found in the LMC only in few young associations/clusters: in the region of SN 1987 (Panagia et al 2000), in the young double cluster NGC~1850 (Gilmozzi et al  1994), in 30~Dor region  (Brandner et al 2001, Romaniello et al 2006), in LH~95 (Gouliermis et al 2002) and in LH~52(Gouliermis et al 2007).

HST ACS/WFC archive photometry in the region of N~11 shows the presence of pre-main sequence candidates (Vallenari \& Chiosi 2007).
The studied fields are  partially covering the associations LH~9, LH~10, and LH~13.
The most conspicuous concentrations of pre-main sequence stars are seen in N11B associated with LH~10, and in LH~9. These concentrations are corresponding to the location of OB stars. A small  group of PMS candidates are found in N11C (associated with LH~13).
 Spectroscopic determination of the age of LH~13 are in the range 3-5 Myr, while LH~10 is 3$\pm 1$ Myr old (Heydari-Malayeri et al 2000). The presence of OB stars in LH~9 set its age at 7.0 $\pm 1 $ Myr (Mokiem et al 2007). Fig.\ref{li_vhel} shows the CMD of the cluster HD~3228 located at the periphery of the association LH~9, where the PMS star candidates are clearly visible as a sequence well separated from the field population. 
The age of this cluster is set to about 3-4 Myr by Walborn et al (1999) based on the presence of a WC star. Pre-main sequence tracks by Siess et al (2000) suggest an age ranging from 1 to 10 Mys, depending on the adopted extinction (A$_V$=0.2 --0.8).  The comparison with the isochrones  shows that pre-main sequence stars have masses from 1.3 ~M$_\odot$ to 2.0  ~M$_\odot$. 
Far infrared Spitzer Space Telescope IRAC and MIPS observations of the region are made in the framework of the SAGE project (Meixner et al 2006). 
These observations are complementary to HST photometry: while HST observations can give information about faint, exposed pre-main sequence candidates, near-IR data allow to detect embedded young stellar objects(YSO).
Comparing the magnitudes and colors of the objects with photometric models by Robitaille et al (2006), Vallenari \& Chiosi (2007) select young stellar object  candidates. They find out that YSO type I (showing large infalling envelopes) and II (characterized by the presence of an optically thick disk) having ages from 0.1 to 1 Myr  are found at the same location than the candidate PMS stars. While it cannot be excluded that LH~9 triggered the star formation in the surroundings, however the data seems to suggest that
the star formation in the region is a long lasting process where stars from 0.1 to 7 Myr are widely distributed.  This seems to be in  agreement with a turbulent scenario of star formation.

\section{Conclusions}
\label{conclu}
In this paper we discuss the young population of stars and clusters in the Magellanic Clouds.
 The last gravitational interaction between LMC and SMC has triggered a star formation episode in both  the cluster and field stars. We find that the SMC was relatively quiescent at ages older than 6-8 Gyr.
This result suggests that at older ages, the tidal interaction
between the Magellanic Clouds and the Milky Way was not able to
trigger significant star formation events. Field star and cluster formation are not completely correlated, as expected on the basis of theoretical models and different star formation modes are expected to act in the field and in cluster populations.
Young stars are well known to  trigger star formation on local scale by injecting energy in the surrounding interstellar medium.   N~11 region in the LMC is one of the most promising candidate of triggered star formation.
 In this paper we discuss the star formation process in this region. We report on the discovery of a PMS candidates associated with N~11 from HST ACS/WFC photometry. 
The PMS are consistent with an age going from 1 to 10 Myr, although a more precise determination is not possible due to the uncertainties on the interstellar extinction. The comparison with the isochrones  shows that pre-main sequence stars having masses from 1.3 ~M$_\odot$ to 2.0  ~M$_\odot$ are present. 
Spitzer IRAC and MIPS observations of the region, made in the framework of the SAGE project reveals the presence of  young stellar object (YSO) candidates. YSO type I and II having ages $< 1 Myr$ are found at the same location than the candidate PMS stars. The data seems to suggest that
the star formation in the region is a long lasting process where stars from 0.1 to 7 Myr are widely distributed, as expected in a turbulent scenario of star formation.



%




\begin{acknowledgements}

This work was done in collaboration with E. Chiosi, E. Held, A. Moretti, G. Bertelli, L. Rizzi

\end{acknowledgements}

\bibliographystyle{aa}

\end{document}